# Lower Bounds on Rate of Convergence of Matrix Products in All Pairs Shortest Path of Social Network

**Dezhou Shen**
Department of Computer Science
Tsinghua University
Beijing, CN 100084
*sdz15@mails.tsinghua.edu.cn*

## Abstract

With the rapid development of social network applications, social network has become an important medium for people to interact. For the minimum distance computation of all pairs in networks, Alon N[4] proposed an algorithm with matrix multiplication, combining with distance product association law and block matrix multiplication, all pairs shortest path length algorithm on networks has time bound $O(\frac{2n^3}{B} \log n)$. In practical applications, considering the scale-free characteristics of social networks and the precision limitations of floating-point operations on computer hardware, I found that the shortest path algorithm has an improved time bound $O\left(\frac{14n^3}{B}\right)$. Based on the above theory, I propose an all pairs shortest path algorithm that combines sparseness judgment and convergence judgment, leveraging the distance product algorithm with matrix multiplication, distance product association law, block matrix multiplication, scale-free characteristics of social networks, and limitation of floating-point operations on hardware. Testing on a social network dataset with 8508 actors, compared to Alon N algorithm, proposed algorithm has a performance improvement of 39% to 36.2 times on CPU and GPU.

## 1   Introduction

In recent years, driven by Asian film industries such as China and India, the global box office has maintained a steady growth trend, and with the rapid development of social network applications, more and more actors use social network for work communication. Therefore, social network has become an important tool for film distribution, marketing, promotion and an important medium for film actors and fans to interact.

Previous studies have lacked research on actors' social networks; PageRank, an existing node influence algorithm, is not accurate in evaluating social network influence. Social network feature measurement is an important task in social network data mining, some network characteristics, such as betweenness and closeness, need to calculate the distances between nodes. As a result, efficient calculation of the shortest path is one of the key problems in social network measurement.

There was Floyd-Warshall[1-2] algorithm for calculating node distances, Aho A V[3] proved that distance product is homogenous to matrix multiplication, Alon N[4] proposed an algorithm to calculate distance product by matrix multiplication, Zwick U[5] optimized distance product using fast matrix multiplication, and Garbow H N[6] proposed an algorithm for distance product using matrix multiplication with element bound.

Table 1: Largest supporting diameter under precision limits of hardware under different number of social network nodes and network diameter

| nodes | diameter limitation | logarithm of diameter | diameter limitation on 32-bit hardware | diameter limitation on 64-bit hardware |
|---|---|---|---|---|
| $10^0$ | 1.7 | 0.5 | 127.9 | 1024.0 |
| $10^1$ | 3.4 | 1.2 | 37.0 | 296.0 |
| $10^3$ | 7.9 | 2.1 | 12.8 | 102.7 |
| $10^8$ | 19.4 | 3.0 | 4.8 | 38.5 |
| $10^{23}$ | 54.0 | 4.0 | 1.6 | 13.4 |
| $10^{61}$ | 141.5 | 5.0 | 0.6 | 5.1 |
| $10^{167}$ | 385.5 | 6.0 | 0.2 | 1.8 |
| $10^{308}$ | 710.2 | 6.6 | 0.1 | 1.0 |

From time bound of matrix multiplication, Strassen V.[7], Coppersmith-Winograd[8], Andrew Stothers[9], Le Gall F[10] proposed the block matrix multiplication with time bound of $O(n^{2.807})$, $O(n^{2.376})$, $O(n^{2.374})$, $O(n^{2.373})$. Monica D. Lam[11] proposed that the progressive time complexity of matrix multiplication with a cache block size of B×B is $O(\frac{2n^3}{B} + n^2)$, and Goto K[12] proposed a new library, OpenBLAS[1], which has a higher cache hit rate and improved time cost in matrix computation on X86 CPU, compared to modern BLAS libraries, such as Intel MKL v8.1.1, ESSL v4.2.0 and ATLAS v3.7.11.

From point of view of matrix sparseness of network adjacent matrix, for sparse matrix multiplication, SciPy[2] and CuPy[3] use SciPy-sparse and CuPy-cuSparse to achieve efficient sparse matrix multiplication on CPU and GPU. For dense matrix multiplication, NumPy[4], SciPy, CuPy and ND4J[5] use OpenBLAS, cuBLAS and other libraries to achieve efficient block matrix multiplication on CPU and GPU.

Therefore, a feasible method for calculating shortest path of full node pairs is: represents the network by adjacent matrix, then makes all matrix elements exponential, computes matrix multiplication and finally takes logarithm of matrix elements. This procedure is called distance product epoch. Iterate several times to get full node pairs shortest path result.

Further, analysis time complexity of the distance product calculation and iteration of each epoch. For time complexity of distance product, matrix multiplication is the most complex sub-procedure; (1) For distance product optimization, Zwick U proposed distance product with block matrix multiplication, therefore choose proper computation device library with optimized cache for block matrix multiplication. (2) For iteration optimization, because Aho A V proved distance product is isomorphic to the matrix multiplication, and regard to that the distance product association law, optimize the iteration process by last distance product result reuse.

## 2  Dataset and Hardware

From January 1, 2011 to March 31, 2015, collected the film actors' social network following relationship, profile information, and text message on Sina Weibo. In all, collected 8508 actors, 577775 edges and 12526114 Sina Weibo posts.

---

1 OpenBLAS: Supports multiple platforms, is based on GotoBLAS optimized BLAS matrix calculation library.
2 SciPy: An open source algorithm library and math toolkit.
3 CuPy: A NumPy-compatible matrix library accelerated by CUDA.
4 NumPy: A Python-based matrix computing library.
5 ND4J: Open source cross-platform matrix computing library developed with JAVA language.

Table 2: Sparseness-aware hardware dependent cache optimized block matrix multiplication libraries with different programming languages.

| Index | matrix sparseness | matrix multiplication implementation | specific library | hardware | programming language |
|---|---|---|---|---|---|
| 1 | Sparse | Python@Operator | SciPy-sparse | CPU | Python |
| 2 | Sparse | CuPy | CuPy-cuSparse | GPU | Python |
| 3 | Dense | NumPy | NumPy | CPU | Python |
| 4 | Dense | CuPy | CuPy | GPU | Python |
| 5 | Dense | ND4J | ND4J-openBLAS | CPU | Java |
| 6 | Dense | ND4J | ND4J-cuBLAS | GPU | Java |

I used a server with two Intel Xeon E5-2620 v4 CPUs and one NVIDIA GeForce GTX 1080 Ti GPU, in total 32 cores with frequency of 2.1Ghz, 128G system memory and 11G GPU memory in my experiment.

## 3 Comparison on block matrix multiplication implementtation

On the actors' social network, this paper implements Strassen block matrix multiplication, Copper-Winograd block matrix multiplication and OpenBLAS matrix multiplication. On the computer hardware, compared the 8508×8508 matrix multiplication time-cost. The block matrix multiplication based on Strassen and Coppersmith-Winograd takes 6 seconds for one iteration of matrix multiplication, and with OpenBLAS library it takes only 4 seconds. The performance of optimized cache block matrix multiplication, such as OpenBLAS matrix computation library, is higher than Strassen and Copper-Winograd block matrix multiplication algorithm.

Therefore, distance product algorithm proposed by Zwick U with OpenBLAS block matrix multiplication is obviously better than distance product algorithm with Strassen and Coppersmith-Winograd.

## 4 Time bound of distance product algorithm with cache optimized block matrix multiplication

For a fully connected network, the graph has a maximum diameter of $n-1$, so using result reuse method in distance product, that is use the last distance product calculation result, the matrix multiplication is a $\lceil \log_2(n-1) \rceil$.

Albert, R[13] estimated the number of documents on the Internet is N=8×10$^8$, diameter D=18.59, found that social networks, the Internet and other complex networks have scale-free and small world phenomenon, diameter is much smaller than node number. I present the corresponding nodes counts, diameters and logarithm of diameters in Table 1, from the observation there is formula (1):

$$D \approx \log N \ll N \tag{1}$$

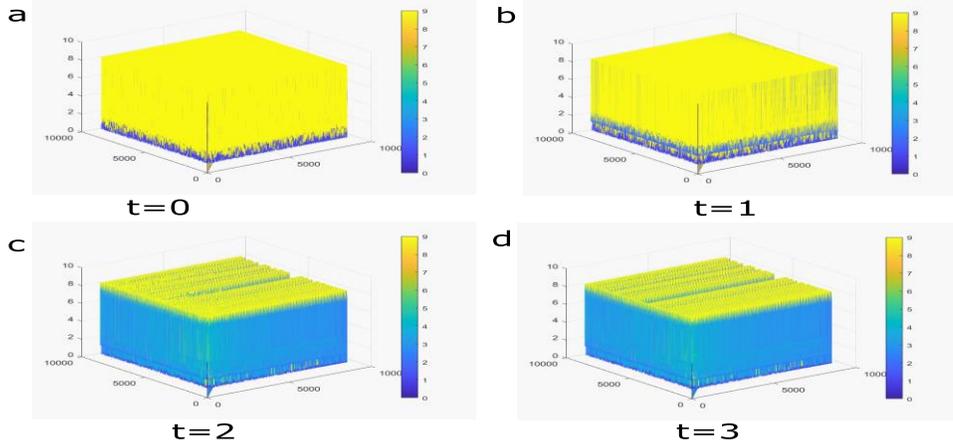

Figure 1: Adjacent matrix, each iterations and the result matrix visualization of actors social network. (a) initial state after each round of calculation, (b) the result matrix after the first round of calculation, (c) the result matrix of the second round of calculation, (d) the third and in the same time last round of calculation, the all pairs nodes shortest path result matrix.

In social networks, the number of nodes N, with cache block size $B \times B$, according to formula (9-10), the matrix product calculation iteration times does not exceed the number of $\lceil \log D \rceil$, recall that Monica D. Lam proved that matrix multiplication time bound is $O(\frac{2n^3}{B} + n^2)$, thus Alon N algorithm time bound is $O\left(\frac{2n^3}{B} \log D\right)$, use the formula (1) result, time bound is $O(\frac{2n^3}{B} \log\log N)$. With the limitation of the floating-point operation on computer hardware, take a network of number of nodes $N < 10^{308}$ as example, as shown in Table 1, see that the shortest path calculation iterations time is no larger than $\log D \approx \log\log N < 7$, that is, the number of calculation iterations does not exceed the magic number 7, bringing a stronger time bound of $O(\frac{14n^3}{B})$. In summary, with the precision limitation of floating-point operations on computer hardware, on scale-free complex networks, all pairs shortest path algorithm time bound is $O(\frac{14n^3}{B})$.

## 5     Shortest Path Algorithm with Block Matrix Multiplication

**Definition 1 Matrix Product** The outer product of two matrices with the same dimension K is the (K×1)×(1×K) matrix multiplication, obtaining (K×K) square matrix, as formula (2):

$$\mathbf{xy}^T = \begin{bmatrix} x_1 \\ x_2 \\ x_3 \end{bmatrix} [y_1, \ y_2, \ y_3] = \begin{bmatrix} x_1 y_1 & x_1 y_2 & x_1 y_3 \\ x_2 y_1 & x_2 y_2 & x_2 y_3 \\ x_3 y_1 & x_3 y_2 & x_3 y_3 \end{bmatrix} \quad (2)$$

**Definition 2 Distance Product** For a matrix A with a dimension of n×m elements, B with a dimension of m×n elements, the distance product of A and B is matrix C with a dimension of n×n elements, for $1 \leq i, j \leq n$, satisfying formula (3):

$$c_{ij} = \min_{k=1}^{m}\{a_{ik} + b_{kj}\} \quad (3)$$

### 5.1     Maximum calculation

Calculate the largest value of the square matrix element $\tilde{x}$, such as the formula (4):

$$\tilde{x} = \max_{i=1}^{n} \max_{j=1}^{n}\{a_{ij}\} \quad (4)$$

Table 3 The characteristics of sparseness, convergence and so on of the intermediate process matrix during the iterative process of the shortest path calculation [a]. For actors' social network with 8508 nodes, calculation process converges after three epochs.

| epochs | maximum element | non-infinite quantity before iteration | non-infinite quantity after iteration | quantity changes | quantity convergence | convergence (%) |
|---|---|---|---|---|---|---|
| 1 | 2 | 617958 | 22627474 | 22009516 | 23435179 | 32.375 |
| 2 | 4 | 22627474 | 71482515 | 48855041 | 72290220 | 99.868 |
| 3 | 8 | 71482515 | 71578359 | 95844 | 72386064 | 100. |
| 4 | 8 | 71578359 | 71578359 | 0 | 72386064 | 100. |

a. Convergence element contains 807705 unreachable edges between nodes.

## 5.2 Exponential calculation

The calculation accuracy of different computer hardware is in line with the IEEE floating-point operation standard[14], first of all, transform the input matrix by exponential process, that is, taking the number of nodes as the base, taking the difference between the largest value of the element $\tilde{x}$ and the matrix elements as exponential, as shown in the formula (5).

$$a'_{ij} = (n+1)^{\tilde{x}-a_{ij}} \qquad (5)$$

As shown in Table 2, for actors' social network with 8508 nodes, the diameter of the network does not allow to exceed 9.8 and 78.4, when calculating a precision of 32-bit floating-point and 64-bit floating-point, otherwise the exponential operation on the computer platform will overflow.

## 5.3 Matrix multiplication

Use different matrix libraries to speed up the matrix multiplication on computer hardware, as shown in Table 3, choose proper device-based optimized matrix multiplication libraries according to different conditions, such as matrix sparseness, computation hardware, and programming language.

## 5.4 Logarithm calculation

Take non-zero elements of the matrix multiplication result to the element logarithm process, that is, the number of network nodes as the base, the element in the matrix multiplication result as the true number, then round down the logarithm result, after that take difference between twice of the largest input matrix element value $\tilde{x}$ and the logarithm result as result, such as formula (6).

$$c_{ij} = 2\tilde{x} - \lfloor \log_{(n+1)} c'_{ij} \rfloor \qquad (6)$$

## 5.5 Distance product association law

From the iteration of the shortest path calculation, suppose A, B, C is the square matrix, according to the association law of distance product, there is a formula (7):

$$A(BC) = (AB)C \qquad (7)$$

Matrix multiplication has association law, and the distance product is isomorphism to matrix multiplication, so simplify the shortest path iteration process as formula (8):

Table 4: Comparison on the performance of shortest path algorithms on different hardware.

| Algorithm | introduction | iterations | time(seconds) |
|---|---|---|---|
| Floyd-Warshall[a] | iterative algorithms imple-mented using Java on the CPU, | - | 1055880. |
| Alon N | the sparse matrix uses CuPy-cuSparse, and the dense matrix uses CuPy calculation on the GPU. | $logn$ | 594.7 |
| PowerLawBound[b] | use NumPy to calculate matrix multiplication on the CPU. | $loglogn$ | 427.9 |
| PowerLawBound[c] | the sparse matrix uses SciPy-sparse, and the dense matrix uses NumPy to calculate on the CPU. | $loglogn$ | 328.4 |
| PowerLawBound[d] | matrix multiplication uses cuBLAS to calculate on the GPU. | $loglogn$ | 95.0 |
| PowerLawBound[e] | matrix multiplication uses openBLAS calculation on the CPU. | $loglogn$ | 45.0 |
| PowerLawBound[f] | matrix multiplication uses CuPy to calculate on the GPU. | $loglogn$ | 19.32 |
| PowerLawBound[g] | the sparse matrix uses CuPy-cuSparse, and the dense matrix uses CuPy calculation on the GPU | $loglogn$ | 15.98 |

a. for reference; b. CPU-NumPy; c. CPU-SciPy-sparse-NumPy; d. GPU-cuBLAS; e. CPU-openBLAS; f. GPU-CuPy; g. GPU-CuPy-cuSparse-CuPy.

$$MM \dots M = (((MM)(MM)) \dots) \qquad (8)$$

Since matrix multiplication has association law and distance product is isomorphism to matrix multiplication, similarly, distance product has an association law. Under the premise of the shortest path matrix $L^{(n-1)}$ given the adjacent matrix and the $n-1$ edge, calculate the shortest path matrix $L^{(n)}$ with $n$ edges, and extend the shortest path of the $n-1$ edge by edge. Calculating the shortest path matrix $L^{(n)}$ can be completed as formula (9):

$$L^{(2)}, L^{(3)}, \dots, L^{(n-1)} \qquad (9)$$

As can be seen from the nature of the shortest path in the graph, the shortest path from the any of two nodes does not exceed $n-1$, so there is formula (10):

$$L^{(n-1)} = L^{(n)} = L^{(n+1)} \qquad (10)$$

**Definition 3 Result Reuse in Distance Product** According to the nature of the formula (8), the process of calculating the shortest path iteration using the last result of distance product follows in turn, as shown in formula (11).

$$L^{(2)}, L^{(4)}, \dots, L^{2^{\lceil \log_2(n-1) \rceil}} \qquad (11)$$

The shortest path needs to be calculated $\lceil \log_2(n-1) \rceil$ times matrix multiplication, from the formula (10), it can be seen that $(n-1) \leq 2^{\lceil \log_2(n-1) \rceil}$ is the same as the simple shortest path matrix calculation result $L^{(n-1)}$, so there is a formula (12):

$$L^{(n-1)} = L^{2^{\lceil \log_2(n-1) \rceil}} \qquad (12)$$

Obviously, the result reuse method saves $n - 2 - \lceil \log_2(n-1) \rceil$ times of matrix multiplication.

Observe the shortest path calculation process of the actors' social network, the adjacent matrix is shown in Figure 1a, and the change of result in each iteration is shown in Figure 1b-d, the shortest path result matrix converges gradually, each element stabilizes as iterations goes.

# 6     Architecture design
## 6.1     Sparseness judgment

As shown in Table 3, observing the adjacent matrix represented by the network diagram is usually a sparse matrix, and becomes a dense matrix during the iteration, and one idea of optimization is to use sparse matrix multiplication to speed up the iteration when the matrix is sparse, and when the matrix becomes denser, the sparse matrix multiplication is much time-consuming.

Taking the actors' social network as an example, the non-zero element of the adjacent matrix of 8508 nodes is 617958, and the non-zero element of the matrix is 0.8536% of the total elements, which is a typical sparse matrix. Consider the advantages of sparse matrix multiplication algorithm, a threshold for triggering sparse matrix multiplication is set, and when the proportion of non-zero-value elements entered is less than the 10% threshold, performs sparse matrix multiplication.

## 6.2     Convergence judgment

For networks with unknown diameters, the calculation results can be made using convergence method, and for the calculation process of reaching convergence, i.e. $L^{(n-1)} = L^{(n)}$, should be regarded as the program termination of iteration, and the calculation of the matrix product method is completed by the formula (4).

## 6.3     The shortest path algorithm that fuses sparseness judgment and convergence judgment

Based on Alon N distance product algorithm, utilize the sparseness judgment of the input matrix and the calculation result convergence judgment, propose a novel shortest path algorithm named PowerLawBound, 'PowerLaw' refers to the network conforming to the power law distribution, 'Bound' refers to the precision of floating-point operation. The name implies two characteristics, one is for the diameter characteristic of scale-free social networks and another is for the limitation precision of floating-point operations on computer hardware, the algorithm process is shown as follows in Algorithm 1.

| Algorithm 1 | Lower Bounds Convergence Matrix Products in All Pairs Shortest Path |
|---|---|
| **Input:** | Adjacent Matrix W, Matrix Row $n$, Network Diameter D |
| **Output:** | Shortest path result $L^{(m)}$ |
| 1: | $L^{(1)} = W$ |
| 2: | m = 1 |
| 3: | **while** m < D: |
| 4: | A = $L^{(m)}$, B = $L^{(m)}$ |
| 5: | For all elements in A and B, calculate bound value $x$ and $y$, so that $x < a_{ij}, b_{ij} < y$ |
| 6: | Exponential function transformation, calculation $A'^{(m)}, B'^{(m)}$ |
| 7: | $a'_{ij} = (n+1)^{y-a_{ij}}$ |
| 8: | $b'_{ij} = (n+1)^{y-b_{ij}}$ |
| 9: | Judging the sparseness of the $L^{(m)}$ and selecting matrix multiplication. |

| | | |
|---|---|---|
| 10: | Matrix multiplication, $C' = A'B'$ | |
| 11: | $c'_{ij} = \sum_{k=1}^{n}(n+1)^{2y-(a_{ik}+b_{kj})}$ | |
| 12: | Taking logarithm of elements in matrix $C'$, the result as matrix $C$ | |
| 13: | $c_{ij} = 2y - \lfloor \log_{(n+1)} c'_{ij} \rfloor$ | |
| 14: | Judging the input and the result convergence, **if** $L^{(m)} = C$: | |
| 15: | **return** $L^{(m)}$ | |
| 16: | $L^{(2m)} = C$ | |
| 17: | m = 2m | |
| 18: | **return** $L^{(m)}$ | |

## 7 Experiment

Firstly, compared to the Floyd-Warshall algorithm implemented on CPU, the Alon N algorithm implementation with matrix multiplication of block cache optimization on GPU has a time performance improvement, see Table 4.

Secondly, compared to Alon N algorithm with matrix multiplication implementation using Strassen, Coppersmith-Winograd and BLAS implementation with block cache optimization such as OpenBLAS, the result shows the cache optimized implementation, OpenBLAS, has more advantages in time performance than Strassen and Coppersmith-Winograd algorithm.

Lastly, compared to the Alon N algorithm of block matrix multiplication with cache optimization, performance of the PowerLawBound algorithm improved ranging from 39% to 36.2 times on CPU and GPU, as seen in Table 4.

## 8 Conclusion

This paper discussed the use of matrix multiplication in calculation all pairs shortest path with association law, and the time bound of Alon N distance product algorithm with cache optimized block matrix multiplication is $O(\frac{2n^3}{B}logn)$. Further strengthen the conditions, in social networks with scale-free characteristics, under the precision limit of floating-point operations on hardware, and with the network diameter limitation to avoid floating-point operation overflow in the computation processes, the algorithm has a lower time bound $O(\frac{14n^3}{B})$. Considered all above limitations, this paper proposed a novel shortest path algorithm, named PowerLawBound, combining matrix sparseness judgment and computation result convergence judgment. 'PowerLaw' refers to the distribution of node number frequencies in networks conforming to the power law distribution, and 'Bound' refers to the precision of floating-point operations on hardware has limitation. The experimental results show that, compared to the Alon N algorithm, PowerLawBound algorithm improves time performance by 39% to 36.2 times on CPU and GPU.

## Broader Impact

This algorithm will change the efficiency of the shortest path calculation and bring positive influence on the graph network representation.